\documentclass[12pt,epsf]{article}

\def\bk{{\bf k}}

\def\bs{{\bf s}}
\def\bi{{\bf i}}
\def\bj{{\bf j}}
\def\bk{{\bf k}}

\def\bS{{\bf S}}

\def\bB{{\bf B}}

\def\bA{{\bf A}}

\def\bO{{\bf O}}

\def\bC{{\bf C}}
\newcommand{\be}{\begin{equation}}
\newcommand{\ee}{\end{equation}}
\newcommand{\bea}{\begin{eqnarray}}
\newcommand{\eea}{\end{eqnarray}}

\newcommand{\lan}{\langle}
\newcommand{\ran}{\rangle}

\usepackage[dvips]{graphics}

\begin{document}

\vskip 2cm
\begin{center}
\Large
{\bf  Contextual Value-definiteness } \\
{\bf and the }\\
{\bf Kochen-Specker Paradox } \\
\vskip 0.5cm
\large
Ken Williams \\
{\small \sl Department of Physics } \\
{\small \sl University of Wisconsin - Parkside } \\
\end{center}
\thispagestyle{empty}
\vskip 0.7cm

\begin{abstract}

Compatibility between the realist tenants of value-definiteness and
causality is called into question by several realism impossibility proofs in
which their formal elements are shown to conflict. We review how this comes
about in the Kochen-Specker and von Neumann proofs and point out a
connection between their key assumptions: a constraint on realist causality
via additivity in the latter proof, noncontextuality in the former. We
conclude that value-definiteness and contextuality are indeed not mutually
exclusive.

\end{abstract}

\newpage

\section{overview}

In contrast to Bell's theorem which draws a contradiction between the
predictions of quantum mechanics and realism, the theorem of Kochen and
Specker (KS), "the second important no-go theory against hidden variable
theories", rather calls into question the very logic in realist thinking.
The argument is directed against a brand of realism characterized by
value-definiteness and noncontextuality, formulated here in section 6.1 as
propositions p(1) and p(2), respectively. When these are applied to an
elementary QM description of spin-1 particle measurements, a contradiction
known as the KS paradox follows.

\noindent In the concluding view of this article, proposition p(2) does not
follow from any realist principle and is even anti-realist in the undue
constraint it places on causality. Moreover, no realist to the writer's
knowledge lays claim to it. As this question concerns the views of realists,
it might well be put to rest with a few excerpts from the writings of
well-known realist thinkers. But that would fall short of our broader aim to
explain how the misunderstanding of a realist p(2) could have come about,
and thus point the way to a reasoned resolution. To this end, we consider
the two principal interpretations of quantum mechanics, orthodox and
realist, with special attention to their contrasting physical
interpretations of the QM wave-function; our approach to the KS paradox is
within this context.

\noindent We also take the view that the paradox may be well understood from
basic quantum mechanics, and so first set about reminding the reader of the
relevant elements that he or she may already know from an elementary study
of the subject. In section 2 we motivate the discussion of discrete
possessed values by considering the quantum signature Stern-Gerlach
experimental data, which, in section 3, is described within the framework of
quantum mechanics. There, the operational equations and constraints
necessary to the KS paradox first appear, equations (\ref{new}). We then
consider in section 4 features of the orthodox interpretation relevant to
the evaluation of the equations: In 4.1 the orthodox identification of the
QM wave-function with individual particle states from which follows in 4.2
its view with regard to individual particle possessed-values. Section 5
deals with the realist interpretation, there too, particularly regarding
features relevant to evaluation of equations (\ref{new}). At the heart of
the interpretation is the
realist identification of the QM wave function with collections of
particles, particle ensembles.

\noindent The KS paradox is derived in section 6 from propositions p(1) and
p(2) applied to the set  (\ref{new}). The  connection to the older von
Neumann impossibility proof is established in 6.2 and to the later work of
Gleason in 6.3. We reconsider the KS paradox in section 7 by close
examination of premise p(2). In the end we find little justification for
proposition p(2) as a realist tenet, and indeed from the words of realist
thinkers, evidence to the contrary. We conclude with an attempt to explain
the natural simplicity of contextuality in realist thinking.

\section{Stern-Gerlach Data}

From the data of Stern-Gerlach (SG) measurements comes a persuasive case
that only certain discrete values of intrinsic angular momentum, spin, are
observable. There, a stream of unprepared electrons enter and exit a small
region of intense inhomogeneous magnetic field strength where it divides in
two, half the electrons deflecting up, the other half down by the same set
magnitude \cite{101} proportional to a given electron's spin projection
along the SG apparatus symmetry axis. Thus measured, each spin projection is
found to have one of the two values, $ \pm \frac{1}{2}$ , for which reason
the electron is called a spin- $ \frac{1}{2}$ particle. A second measurement
along the same SG axis taken on either of the two sub-ensembles invariably
confirm the previous result: electrons previously deflected up (down) due to
positive (negative) spin of magnitude $ \frac{1}{2}$ along the axis, are
again deflected up (down)
\begin{center}
\scalebox{1.5}[1.5]{\includegraphics{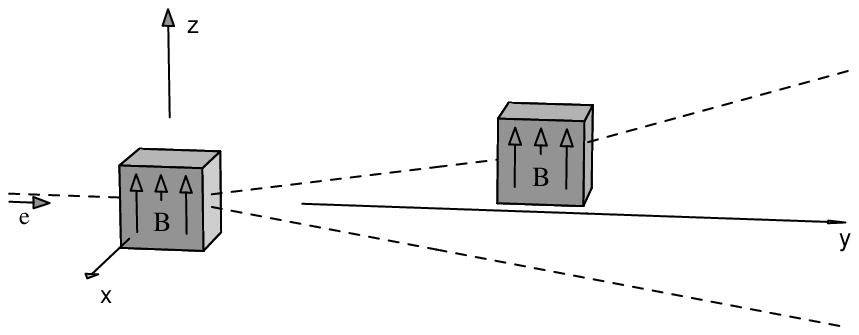}}\\
figure 1
\end{center}
where $\hat{\bB } $ in figure 1 gives the averaged SG B-field direction. As
the pre-measured spin orientations are assumed random, isotropically
distributed and uncorrelated to the orientation of the measuring apparatus,
this phenomena is unexpected. One would from classical considerations
expect the deflection magnitudes to vary continuously with the
relative spin-to-apparatus angle
\bea
d  \propto s_z & = & s \cos \phi
\eea
having maximum and minimum values for $ \phi = 0  \&  \frac{ \pi }{2} $,
respectively
\begin{center}
\scalebox{1.5}[1.5]{\includegraphics{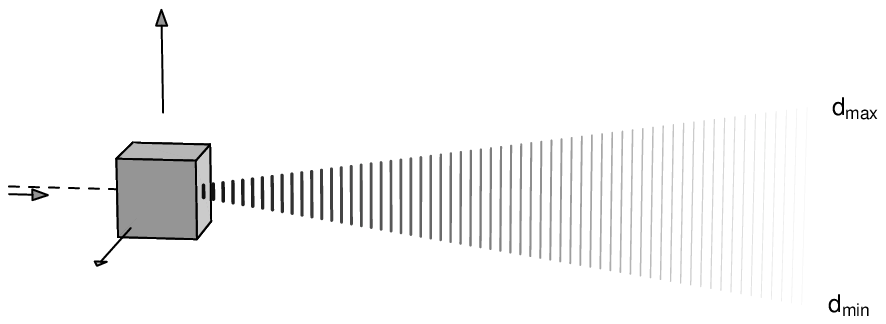}}\\
figure 2
\end{center}
One finds however that when the apparatus itself is re-oriented (to test
against apparatus-to-ensemble correlation), the same result obtains
\begin{center}
\scalebox{1.5}[1.5]{\includegraphics{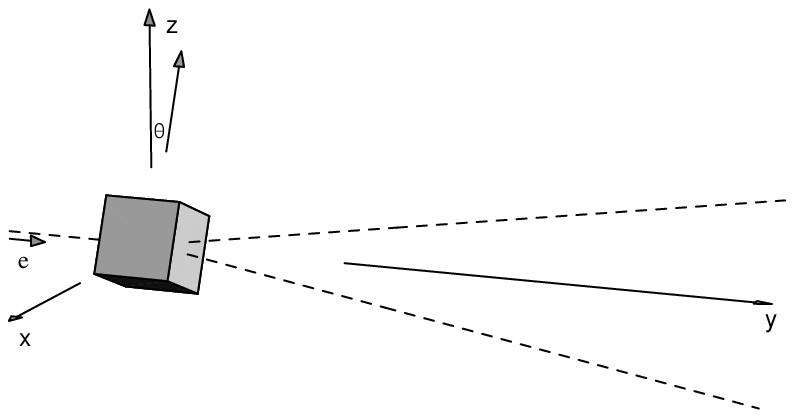}}\\
figure 3
\end{center}
half electrons deflecting up again by the same magnitude as before, half
down, only
now along the re-oriented axis. Moreover, when this second measurement is
taken
along an axis perpendicular to the first, again, half are deflected up, half
down, by the same magnitude. Indeed, one finds that no matter the SG
orientation or the order in which one appears in a sequence of such
measurements, the results are always $ + \frac{1}{2} $ or $ - \frac{1}{2}$
only . The electron's spin projection is therefore said to be quantized
(i.e.,  given in discrete, set amounts) along the measurement axis, called
then the axis of quantization, in the figure below the SG
symmetry axis specified by $ \theta $ and $\theta^\prime$, SG orientations
relative to the laboratory z-axis.
\begin{center}
\scalebox{1.5}[1.5]{\includegraphics{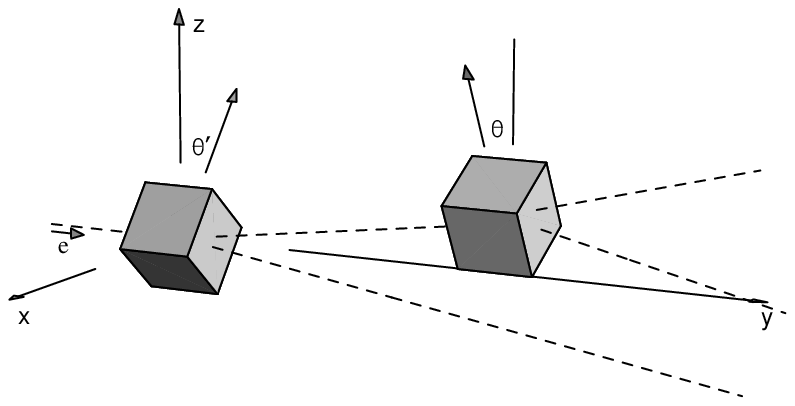}}\\
figure 4
\end{center}
The phenomena poses a difficulty to ordinary intuition. Imagine e.g. that
whenever you looked up at the moon its polar tilt (as measured from the
plane of sight) varied in
$ 180^o $ increments only
\begin{center}
\scalebox{0.7}[0.7]{\includegraphics{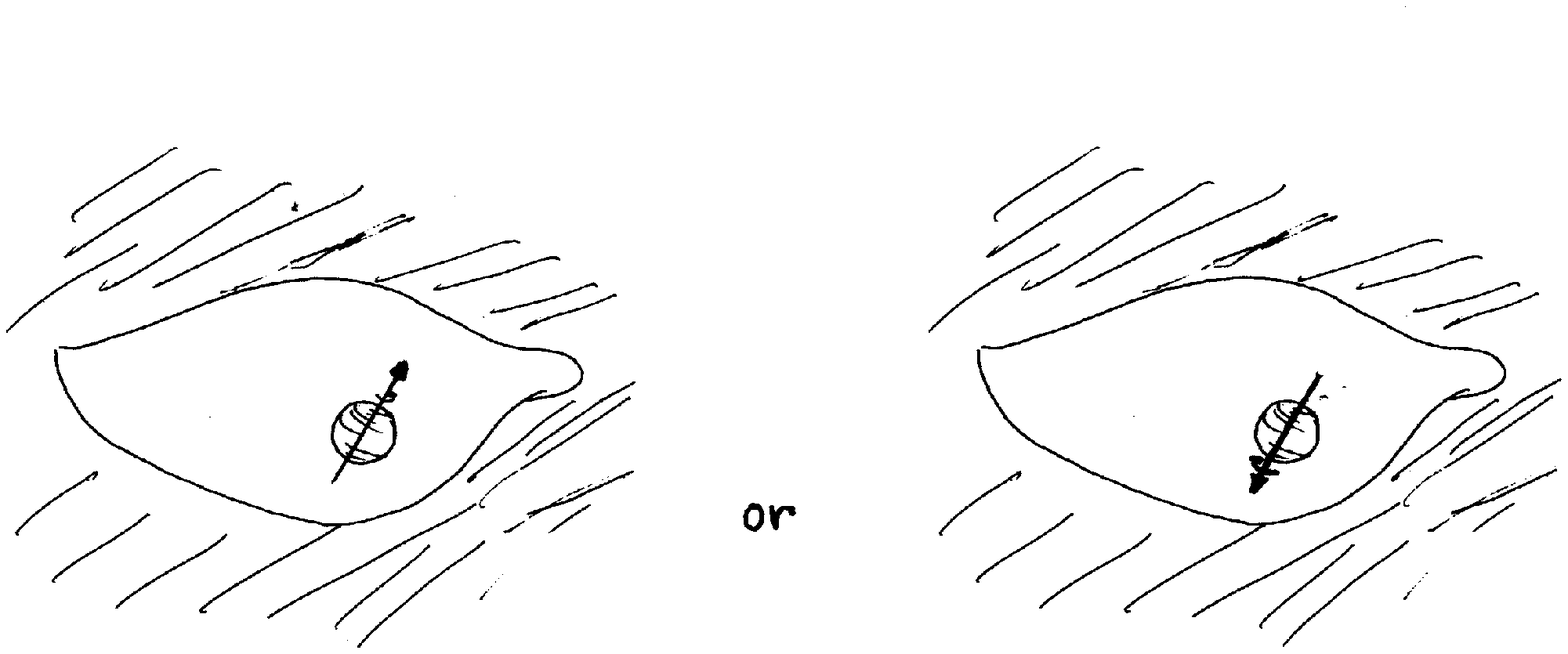}}\\
\end{center}
no matter the orientation of your head (as measured from the local earth
plane).
\begin{center}
\scalebox{0.7}[0.7]{\includegraphics{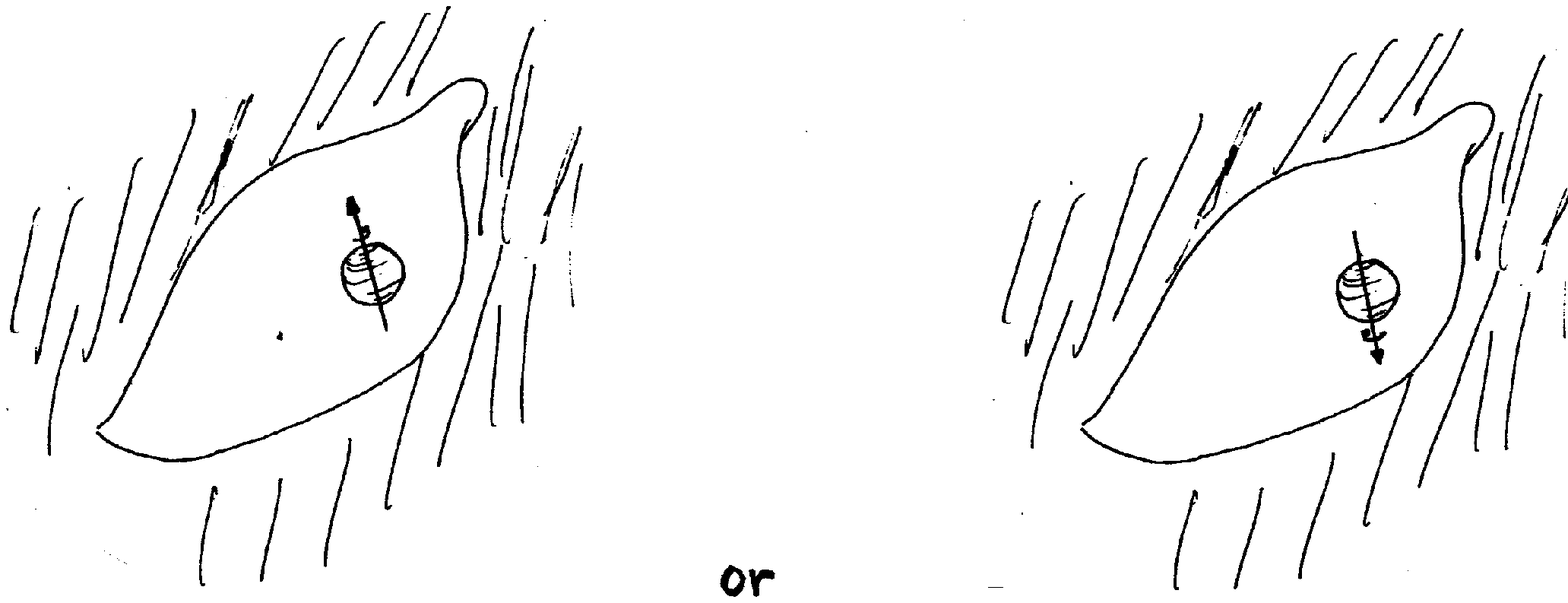}}\\
figure 5
\end{center}
The moon's angular momentum would then be said to be quantized.... For the
electron this and other unusual
microscopic phenomena are well described by the Physics of small-scales,
quantum mechanics.

\section{quantum mechanical description}

Relative frequencies for each of the spin-up, spin-down classes of electron
data described above may be obtained from the formalism of quantum
mechanics. Experimentally, the frequencies are statistical probabilities
over ensembles, collections of identical elements, strictly valid in the
large ensemble limit. As regards individual measurements, however, the
theory makes no prediction \footnote{ See e.g. ref. \cite[p. \ 31]{103} and
ref. \cite[p. \ 2]{102} }. The up (down) deflection probability at apparatus
$ \theta $ for a sub-ensemble of electrons previously deflected up at
apparatus $ \theta^\prime $ is given by $ \cos^2 \frac{ \phi }{2} $ ( $
\sin^2 \frac{
\phi }{2} $ ), $ \phi \equiv \theta - \theta^\prime $, assuming for
simplicity SG rotation with respect to the laboratory y-axis.
\begin{center}
\scalebox{1.5}[1.5]{\includegraphics{fig2mc.ps}}\\
figure 6
\end{center}
These probabilities are expectation values of projector
operators $ {\rm P }_{ \theta \pm} $  in a complex Hilbert space of state
vectors $ | \Psi \ran $ whose calculation proceeds as follows:
\begin{equation}
\lan  {\rm P }_{ \theta \pm} \ran =  _{\theta^\prime}\lan + | {\rm P }_{
\theta \pm}| + \ran_{\theta^\prime} = \left\{ \begin{array}{l} \cos^2 \frac{
\phi }{2}     \\  \sin^2 \frac{ \phi }{2} \end{array} \right.  \label{1}
\end{equation}
where
\bea
|  + \ran_\theta  &=& \cos\frac{\theta}{2} \, | + \ran +  \sin\frac{ \theta
}{2} \, | - \ran  \\
{\rm P }_{ \theta \pm}  &=&  | \pm \ran_{ \theta }  \, _{\theta }\lan \pm |
\label{nnew}
\eea
with relations
\bea
\lan \pm | \pm \ran & = & 1 \nonumber \\
\lan \pm | \mp \ran & = & 0 . \label{ttwo}
\eea
For deflected-down prepared subensemble predictions, vector $ | +
\ran_{\theta^\prime}  $ in (\ref{1}) is replaced by $ |  - \ran_{
\theta^\prime } =  \cos\frac{ \theta^\prime }{2} | - \ran -  \sin\frac{
\theta^\prime }{2} | + \ran  $.

\noindent Spin-up/spin-down vectors, $ | \pm \ran $ , are so-called by
analogy with the relation between ordinary classical spin vectors, $ \bs_\pm
$, and what would be their SG deflections. However in the
two-dimensional Hilbert space H2 which they span, the directions of $ | \pm
\ran $  are not antiparallel, but mutually perpendicular, corresponding to
dual, mutually exclusive experimental outcomes.
\begin{center}
\scalebox{1.5}[1.5]{\includegraphics{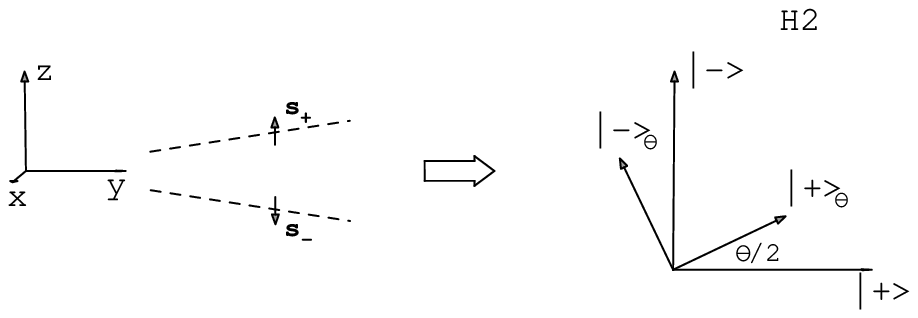}}\\
figure 7
\end{center}
Given projectors $ {\rm P }_{\theta \pm} $ , both vectors are completely
specified by relations
\bea
{\rm P }_{\theta \pm } | \pm \ran_\theta & = & (1) \cdot | \pm \ran_\theta =
| \pm \ran_\theta  \nonumber \\
{\rm P }_{\theta \pm } | \mp \ran_\theta & = & (0) \cdot | \mp \ran_\theta =
0
\label{2}
\eea
called eigenvalue (proper- value) relations, where (\ref{2}) states that $ |
+ \ran_\theta $ is an eigenvector of both projector operators $ {\rm P
}_{\theta + }$ and $ {\rm P }_{\theta - }$  having eigenvalues 1 and 0,
respectively; these are sometimes called "yes" and "no" eigenvalues, and
projectors, accordingly, yes-no operators. In consequence, $ | \pm
\ran_\theta
$ are also eigenvectors of the quantum mechanical spin operator, $ S_\theta
= \frac{1}{2} ( {\rm P }_{\theta + }- {\rm P }_{\theta - })  $ , having
eigenvalues $ \pm \frac{1}{2} $
\bea
S_\theta | \pm \ran_\theta & = & \pm \frac{1}{2} | \pm \ran_\theta
\label{3}
\eea
which gives immediately that the vectors $ | \pm \ran_{ \theta^\prime } $
with $ \theta^\prime \neq \theta  $  are generally not spin $ S_\theta $
eigenvectors.

\noindent Vectors and operators may also be expressed in 2-dimentional
matrix form
\begin{equation}
| + \ran_\theta =  \left( \begin{array}{r}\cos\frac{\theta }{2} \\
\sin\frac{ \theta }{2}  \end{array} \right) \qquad \qquad \qquad \qquad
| - \ran_\theta =  \left( \begin{array}{ r }- \sin\frac{\theta }{2} \\
\cos\frac{ \theta }{2}  \end{array} \right) \label{2-vecs}
\end{equation}
\begin{equation}
{\rm P }_{ \theta + } =  \left( \begin{array}{ r r }\cos^{2}\frac{\theta
}{2}& \frac{1}{2} \sin\theta \\ \frac{1}{2}\sin\theta & \sin^{2}\frac{
\theta }{2}  \end{array} \right) \qquad \qquad
{\rm P }_{ \theta - } =  \left( \begin{array}{ r r }\sin^{2}\frac{\theta
}{2}& - \frac{1}{2}\sin\theta \\ - \frac{1}{2}\sin\theta & \cos^{2}\frac{
\theta }{2}  \end{array} \right)
\end{equation}
whereby relations (\ref{1}), (\ref{ttwo}), and (\ref{2}) above
are of course preserved. In addition, for each quantization $ \theta $  we
have the
projector relations
\begin{equation}
{\rm P }_{\theta + } + {\rm P }_{\theta - } = \left( \begin{array}{c c} 1 &
0 \\ 0 & 1 \end{array} \right) \equiv { \bf 1 }
\end{equation}
\bea
{\rm P }^2_{\theta \pm } & = & {\rm P }_{\theta \pm }
\eea
known respectively as completion and projector conditions. Their symbolic
evaluation over the ensemble yields the expectation value equation
\bea
\lan {\rm P }_{\theta + } \ran_\theta + \lan {\rm P }_{\theta - }
\ran_\theta & = & 1
\eea
with constraint
\bea
\lan {\rm P }_{\theta \pm } \ran_\theta & = & 1 \, {\rm or } \, 0 \, .
\eea
It is sometimes of interest to compare sets of eigenvector pairs for
selected quantizations. From (\ref{2-vecs}) we tabulate for orientations
$\{ \theta \} = \{0,\pi/2, \pi, 3\pi/2 \} $.
\begin{center}
\scalebox{0.9}[0.9]{\includegraphics{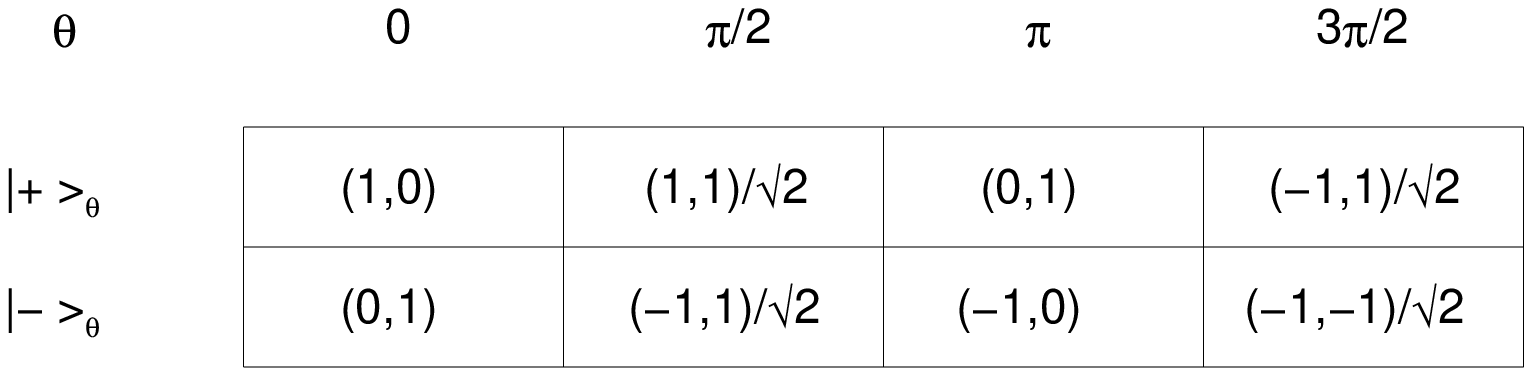}}\\
table 1
\end{center}
Notice that eigenstates are not necessarily exclusive to single
orientations: $ |- \ran_{\pi/2} = |+ \ran_{3 \pi/2} $, so that e.g. from
(\ref{nnew}), $ {\rm P }_{\pi/2_-} = {\rm P }_{3 \pi/2 + } $. As the
ambiguity is significant to the coming discussion, for emphasis we now make
a notational change. We specify a Hilbert space eigenstate now by the
spatial direction of its classical counterpart, one related to the other as
illustrated in
figure (7). Then, $ | \pm \ran_{\theta^\prime } \sim | \bs^\prime_\pm \ran
$. $ {\rm P }^\prime_i$  thus projects out the QM state with classical
counterpart $ \bs^\prime_i $ measured by an SG apparatus with orientation $
\theta^\prime $, which, interchangeably with the corresponding average SG
field direction $ \hat{\bB^\prime } $, we shall call the measurement's {\it
context}. The two notations are related by
\bea
\hat{\bB}^\prime \cdot \hat{\bk} &=& \cos(\theta^\prime ) \nonumber \\
\hat{\bs}^\prime_{1,2} & \equiv & \hat{\bs}^\prime_{+,-} = +,-
\hat{\bB}^\prime \nonumber \\
{\rm P }_{1,2}^\prime &=& {\rm P }_{\theta^\prime +,- }
\eea
We therefore have from table 1
\bea
{\rm P }_2^{(\pi/2)} &=& {\rm P }_1^{(3\pi/2)}
\eea
consequent upon $ \bs_2^{(\pi/2)}= \bs_1^{(3\pi/2)} ( = \hat{\bi }) $: The
spin-down SG outcome state for SG orientation $ \theta = \pi/2 $ is the same
as the spin-up outcome state for SG orientation $ \theta = 3\pi/2 $   . In
the new notation table 1 becomes
\begin{center}
\scalebox{0.9}[0.9]{\includegraphics{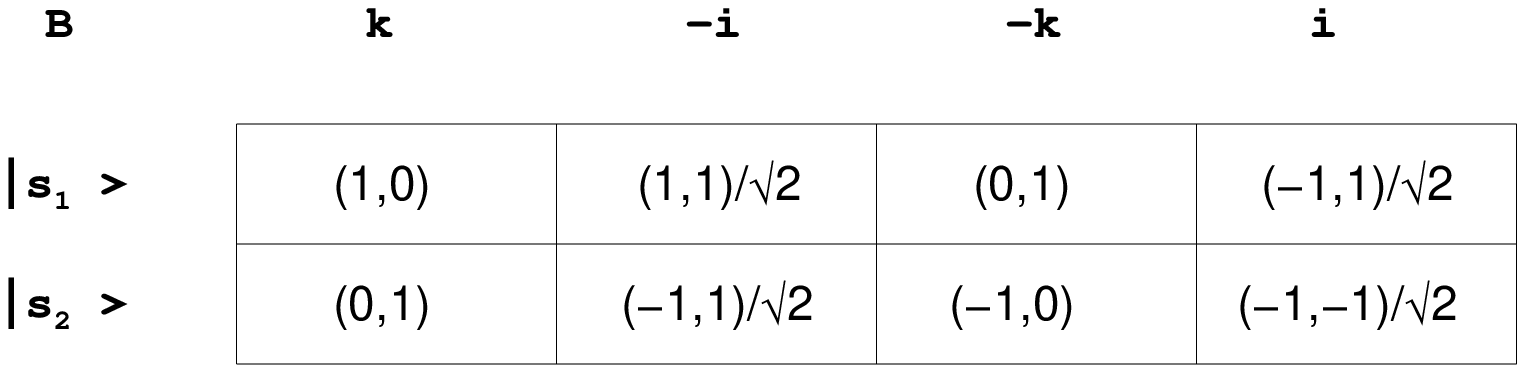}}\\
table 2
\end{center}
For the purposes of the KS analysis here we will need to consider the
structure of measurement outcomes on spin-1 particles, measurements also
made by the SG apparatus. The observed mutually exclusive outcomes are:
deflection up, deflection down,
and no deflection at all.
\begin{center}
\scalebox{1.5}[1.5]{\includegraphics{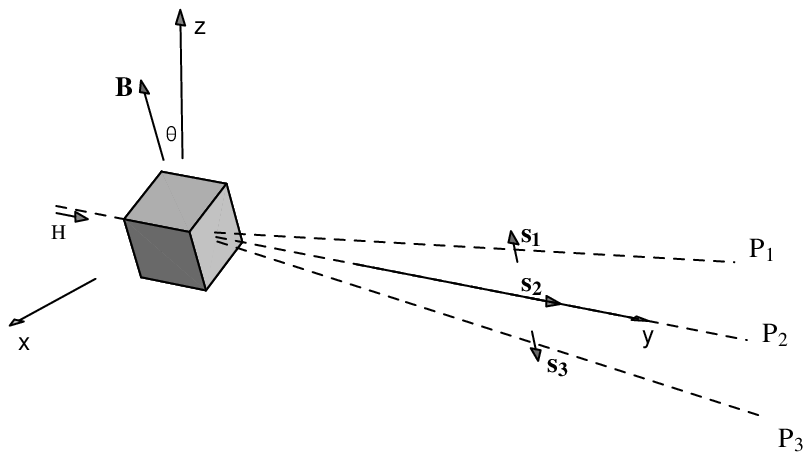}}\\
figure 8
\end{center}
In the now corresponding  {\it three} dimensional Hilbert space of states we
have operator
relations
\begin{equation}
{\rm P }_1  + {\rm P }_2  + {\rm P }_3 = \left( \begin{array}{c c c} 1 & 0 &
0 \\ 0 & 1 & 0 \\ 0 & 0 & 1 \end{array} \right) \equiv { \bf 1 }
\end{equation}
\bea
{\rm P }^2_i  & = & {\rm P }_i
\eea
and related expectation-value completion and projection relations
\bea
\lan {\rm P }_1 \ran^\prime + \lan {\rm P }_2  \ran^\prime  + \lan {\rm P
}_3 \ran^\prime  & = & 1 \nonumber \\
\lan {\rm P }_i  \ran^\prime & = & 1 \, \, {\rm or } \, \, 0
\qquad \qquad \qquad  {\rm for } \; \bB^\prime = \bB
\label{new}
\eea
a constrained sum rule (csr) central in the derivation of the KS paradox.
Looking ahead, we now tabulate the eigenvector triplets for the four
contexts $ \{ \bB \} =
\{ \hat{\bk},- \hat{\bi},- \hat{\bk},\hat{\bi} \} $
\begin{center}
\scalebox{0.9}[0.9]{\includegraphics{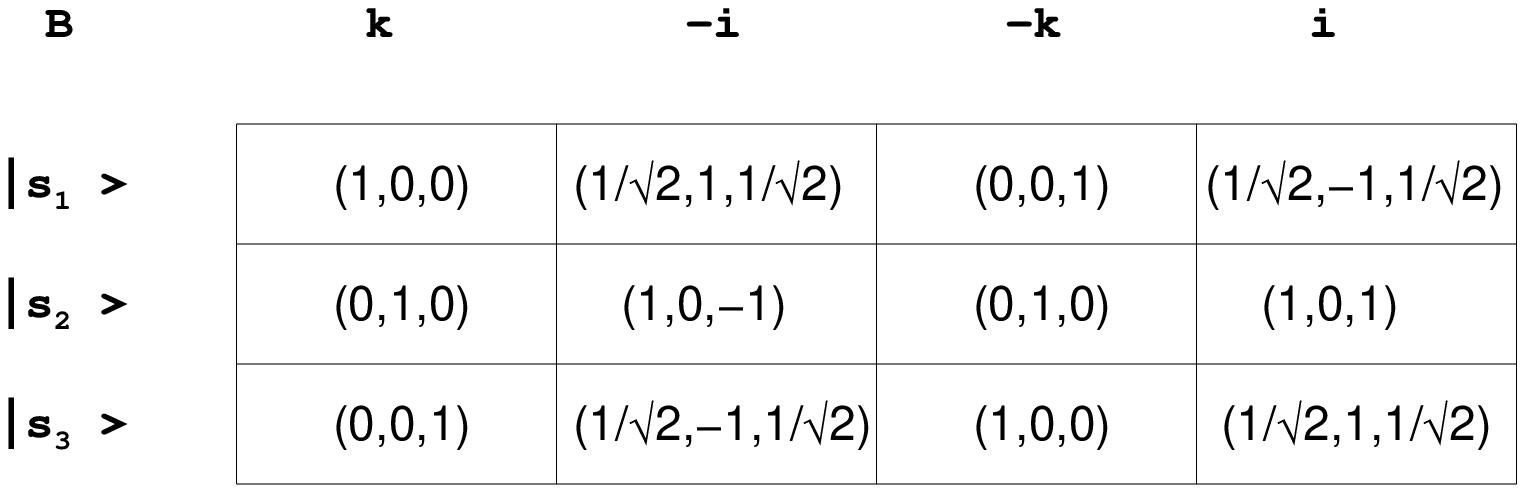}}\\
table 3
\end{center}
Table 3 and csr (\ref{new}) taken together lend themselves to various
physical
interpretations, most of which may be grouped into one of two main
categories: orthodox and realist. The KS contention detailed in section 6 is
that an expanded  table 3 (obtained simply by expanding the number of
contexts considered) is not amenable to any self-consistant realist
interpretation; this is the KS
paradox. For this reason, we begin in the next two sections by considering
such features of the two broad interpretations as are relevant to the KS
discussion.

\section{Orthodox interpretation of SG data and QM
description}

\subsection{ Idea of the individual electron state }

There are several ways in which the behavior of individual electrons under
the influence of SG measurements is similar to that of quantum-mechanical
spin eigenfunctions under the action of projector operators. As a prepared
electron deflects either up or down along a measuring SG axis, a projector
operator will project an arbitrary state $ \Psi $  into either a spin-up or
a spin-down eigenstate characteristic of the SG orientation $ \hat{\bB} $
\footnote{ The actual projection on an arbitrary state is by prescription
only; a state is said to "collapse" upon measurement to one of the context's
eigenstates, though no entirely satisfactory dynamics for the collapse has
ever been worked out \cite{125}. The shortcoming is closely related to the
so-called measurement problem in QM. }.
\bea
{\rm P } | \Psi \ran &  & \stackrel{\longrightarrow }{collapse} \quad |
\bs_1 \ran \quad or \quad | \bs_2  \ran
\eea
An electron spin prepared along a given measurement axis then will not be
altered upon measurement - a deflected up(down) prepared electron deflects
up(down) again - just as projector operators project their own eigenstates
onto themselves. On the other hand, if the prepared and measurement axis are
not identical, the electron deflects up or down with probability amplitude
given by the corresponding state projected eigenfunction coefficient
\begin{equation}
{\rm P }_{1,2}^\prime | \bs_1 \ran  = \left\{ \begin{array}{l} \cos\frac{
\phi }{2}| \bs_1^\prime \ran   \\  \sin\frac{ \phi }{2}| \bs_2^\prime \ran
\end{array} \right.  \label{10}
\end{equation}
illustrated
\begin{center}
\scalebox{1.5}[1.5]{\includegraphics{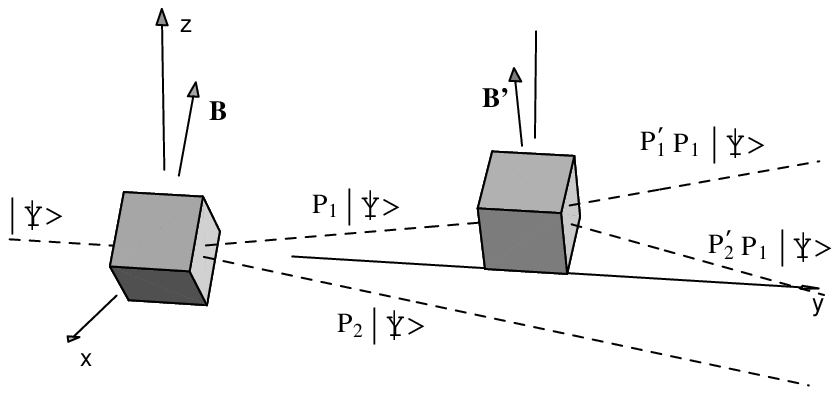}}\\
figure 9
\end{center}
And so one might well identify the quantum mechanical spin- $ \frac{1}{2} $
wave function $|\Psi \ran $ with individual electron states, projector
operators with the measuring apparatus, and action of the operator on the
wave function with the measuring process, just the
identifications in fact made in the conventional or orthodox interpretation
of QM \footnote{ See e.g. ref. \cite[pp. \ 108 \& 214]{101} .}. While the
assignments are largely unimportant to the practical workings of the
formalism, the calculation of expectation values, they offer a picture of
the world of individual electrons, a kind of metaphysics.

\noindent One difficulty with the interpretation is that spin-vectors are
particular to their quantization, having definite projections only along the
quantized directions of their measurement axis; in all other directions the
projection is not well-defined. The same then in this interpretation for the
electron spin itself; the orthodox electron does not possess a
spin-projection along a given direction except upon measurement (and prior
to subsequent distinct measurements); unlike ordinary classical spin
projections, an electron's seems to
exist only along one set of directions at a time.

\noindent Another difficulty is that "identical" electrons - those of a
given ensemble and so represented by the same spin function - generally
exhibit different behaviors when subjected to the same conditions: Some
electrons of a deflected-up subensemble e.g. will deflect up, others down,
when their spin projections are measured along a direction different from
the preparation axis. The phenomena is at odds with the principle of the
identity of indiscernibles \cite{122} (see also discussion in \cite[p. \
8]{116}, and "sufficient reason" principle in \cite[p. \ 266]{104}  ).

\noindent Finally, the probabilities of QM have meaning to the
experimentalist only in reference to ensemble measurements. \footnote{ see
ref. \cite[p. \ 2]{102} and \cite[p. \ 36]{103} }

\subsection{ Orthodox interpretation of csr (\ref{new})  }

The value of the expectation  $ \lan  {\rm P }_i  \ran_\Psi  $ answers the
question whether a particle in state $ \Psi $ has a definite spin in context
$ \bB $ : 1 or 0 for yes, no otherwise. From (\ref{new}) we see that the
answer is "yes" generally when the particle has been prepared within the
measurement context of interest, $ |\Psi \ran \to |\bs_j^\prime \ran $ with
$ \bB^\prime = \bB $. But the answer might also be yes for a different
preparation context,  $ \bB^\prime \neq \bB $, provided the classical
companions to Hilbert space vectors $ | \bs_j^\prime \ran  $ and $ | \bs_i
\ran  $, $ \bs_j^\prime \, \& \, \bs_i $, coincide for some i and j. We
illustrate such an instance
\begin{center}
\scalebox{1.5}[1.5]{\includegraphics{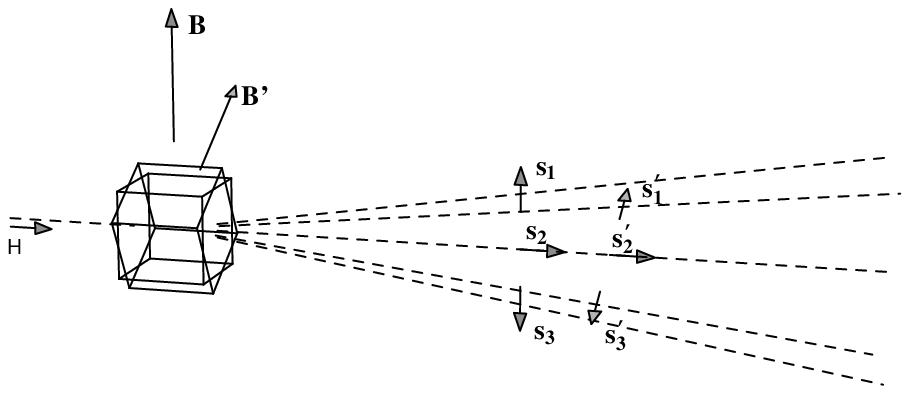}}\\
figure 10
\end{center}
In this case the projection measurement with result corresponding to that
for classical spin direction $ \bs_2 = \bs_2^\prime $ can be made within
either context $ \bB $ or $ \bB^\prime $ with equal results, and is
therefore said to be context independent (wrt contexts $ \bB $ and $
\bB^\prime $ ), or noncontextual. A prepared particle $ | \bs_j^\prime \ran
$
might therefore have a definite spin in all contexts with quantization $
\bs_j^\prime $  and can be said to {\it possess} the corresponding set of
projector operator eigenvalues. In all other contexts its spin projections
are not definite, and the particle could not be said to possess the
corresponding projector eigenvalues.

It is easy to see that the number of noncontextuality constraints imposed by
this interpretation on a set of sum rules (\ref{new}) corresponding to a
large number of contexts considered simultaneously would be relatively
small. Not so in the realist interpretation, as we shall see in the next
section. It is this consequence of the realist interpretation that the KS
analysis in the section following exploits to draw a self-contradiction.

\section{Realist interpretation of SG data and QM description }

\subsection{ Deterministically-held possessed values and the realist
extension to csr (\ref{new}) }

In the realist view there is a sense in which an electron has definite spin
projections in all directions at all time; its physical
characteristics combined with preparation determines {\it  unambiguously}
whether an
electron will deflect up or down in a given Stern-Gerlach measurement,
nothing left to chance. "The good lord does not throw dice", is how Einstein
is reported to have put it \cite[p. \ 190]{104}. Then, in the event that
otherwise identical electrons have different deflections their preparations,
their causal histories, must also have been different.

\noindent { \it In this sense } it might be said that an electron is in
possession of a deflection value, "up" or "down", and hence in possession of
its spin projection, prior to measurement. In much the same sense we say
that a massive object possess a weight, referring to the weight value $
\nu(w) $ a scale will read when the object is placed on it.This does not
conflict with the possibility that one scale gives a different reading from
another; one may speak of a body's terrestrial weight, its weight on the
moon,
on mars, etc.. A massive body, like the realist electron, is at all times in
possession of an infinite number of measurement values, the set of values
that (for the body) weight scales may measure for each of the infinitely
many possible measurement situations. These values are deterministically
well-defined and so, given the object's physical characteristics and the
deterministic factors acting upon it, the causal chain leading up to the
measurements, may be identified with that oject. By this identification - in
this sense - we say that the
values are possessed by the object - weights of a body, spin projections
of an electron.

\noindent A body's weight values might of course be determined from its
mass, given the relevant celestial masses and the law of gravity; in a world
of weight measurements however a statement of either is equivalent.
Likewise, in absence of a knowledge of some underlying electron spin and the
dynamics of microscopic spin-measurements, we may meaningfully identify a
set of possible measurement results $ \{ \nu(s) \} $ with an electron.
Fundamental to the realist view is that an object's possessed values are, in
principal, verifiable measurement values.

\noindent The description is compatible with the realist tenet that
\begin{equation}
{\it Distinct \,\, phenomena \,\, supervene \,\, upon \,\, distinct \,\,
causes \,\,
only } \footnote{ cf. opposing accounts of the "realist view" in \cite[p. \
19]{116} and \cite[pp. \ 12-13]{118} }. \label{cause}
\end{equation}
on which is based the EPR reality criterion
\begin{quotation}
\small
\noindent If, without in any way disturbing a system, we can predict with
certainty (i.e., with probability equal to unity) the value of a physical
quantity, then there exists an element of physical reality to this physical
quantity.
\end{quotation}
which set the stage for their famous argument in defense of realism
\cite{105} and lead eventually to the Bell inequality constraint on the
same \cite{5}. In Bell's analysis, the above description of electron
possessed values $ \{ \nu(s) \} $ also appear, depending too upon the manner
in which the spin measurements are taken, their contexts. And it is only
upon stringent space-time constraints that context dependence is ruled out.

\noindent We now use the notation $ \{ \nu(s) \}^\prime $ to denote the set
of electron spin projection values measurable in context $ \bB $, though
possessed by the particle while in context $ \bB^\prime $ . Similarly for
the set of projector (yes-no) values
\bea
\{ \nu({\rm P } ) \}^\prime & \equiv &  \{ \nu^\prime({\rm P }_1 ),
\nu^\prime({\rm P }_2  ), ...  \nu^\prime({\rm P }_n )  \}
\eea
where $ \nu^\prime({\rm P }_i ) $ is the yes-no value for the $ i^{th} $
quantized direction (corresponding projector eigenvalues in descending
order) of context $ \bB $, possessed by a particle while in context $
\bB^\prime $. We identify realist possessed values with QM operators in this
notation only for the sake of comparison with QM expectation values with
which they sometime coincide; the value $ \nu({\rm P } ) $ simply answers
the question whether or not a spin-measured quantum particle would scatter
in the same direction as a classical particle of spin  $ \bs $, and is not
to be understood as a formal mapping of the QM projector  $ {\rm P } $. The
correlation between expected and experimental values is of primary
importance in the realist view, that between expected values and the QM
formalism (or any other) of secondary importance; the extent to which
eigenvalues of a QM operator are observed values is the same to which the
realist possessed value $ \nu(\bO) $ is an eigenvalue
of the QM operator $ \bO $.

\noindent By realist extension of csr (\ref{new}) to all contexts, then, we
have:
\bea
\nu^\prime({\rm P }_1 )+\nu^\prime({\rm P }_2 )+\nu^\prime({\rm P }_3 ) & =
& 1  \nonumber \\
  \nu^\prime({\rm P }_j  ) & = & 1 \, or \, 0 \qquad \qquad \qquad  \forall
\quad \bB^\prime . \label{real eqns}
\eea
Lastly, we may attribute to a  hypothetical realist state function, one
completely describing the state of an individual particle, the complete set
of the particle's possessed values
\bea
\psi &=& \psi [ \{ \{ \nu({\rm P } ) \}^\prime \} ]  \qquad \qquad  \forall
\quad \bB , \; \bB^\prime. \label{g}
\eea
The realist particle thus represented by $ \psi $ has definite spin
projections in all H-spatial directions, simultaneously in all contexts.

\subsection{Relation between realist $ \psi $ and QM $ \Psi $  }

\noindent We symbolically represent the value assignment for an
electron-spin measurement
\bea
\lan \psi | \bS | \psi \ran  &=& \nu(\bs )  \to \pm \frac{1}{2} \, .
\label{c}
\eea
For sets of realist possessed values of mutually non-commuting contexts,
however,
\bea
\left[  \bS_\theta , \bS_{ \theta^{ \prime } } \right]  & \neq & 0  \, ,
\eea
QM predicts and observation confirms that a measurement of one set alters
the elements of the other; determination of such simultaneous possessed 
value sets is therefore not possible.

\noindent We illustrate a possible difficulty to realist thinking that this
resolves by first reverting back briefly to the former notation and consider
the QM operational identity
\bea
\bS_{\pi/4} &=& (\bS_0 + \bS_{\pi/2} )/\surd2 \label{d1}
\eea
with the above realist value assignments (\ref{real eqns}) applied to its QM
expectation
\bea
\lan \Psi | \bS_{\pi/4 } | \Psi \ran &=&  ( \lan \Psi | \bS_{0} | \Psi \ran
+ \lan \Psi | \bS_{\pi/2 } | \Psi \ran )/\surd2  \rightarrow \pm \frac{1}{2}
= (\pm \frac{1}{2} \pm \frac{1}{2})/\surd2 \label{h0}
\eea
a statement manifestly false for all $\pm $ combinations. The mistake
results from an essential difference between the eigenvalues $\nu(\bO) $ and
expectation values $ \lan \Psi | \bO | \Psi \ran $ of an operator $ \bO $.
We refer however to the use of relation (\ref{d1}) for its {\it meaning } 
which lies in the agreement obtained between its
expectation value and the experimental average of measurements made upon
many, many electrons for each of the three indicated contexts, i.e. in an 
agreement
with empirical relations between sub-ensemble averages. In practice, then,
(\ref{d1}) pertains to three entirely disjoint electron sub-ensembles, each
measured within one of the three contexts, $ \{ 0, \pi/4, \pi/2 \}  $, so
that no single electron spin is measured more than once; the meaning of
(\ref{d1}) derives from the identification of quantum mechanical state
functions $ \Psi $ with subensembles \cite{111}, and not from a direct
association with the possessed values of individual ensemble members
described by the realist $ \psi $. With $ | \Psi \ran $ as the QM state
function, the expectation value relation
\bea
\lan \Psi | \bS_{\pi/4 } | \Psi \ran &=&  ( \lan \Psi | \bS_{0} | \Psi \ran
+ \lan \Psi | \bS_{\pi/2 } | \Psi \ran )/\surd2     \label{h1}
\eea
is found to correspond to the observed relation
\bea
\bar{ \bS}_{\pi/4} &=& ( \bar{ \bS}_{0} + \bar{ \bS}_{\pi/2}) /\surd2
\label{h2}
\eea
where
\bea
\bar{\bS}_\theta & \equiv & \frac{1}{2}( n_+ - n_- )/N
\eea
$ n_\pm $  the number of ensemble electrons observed to deflect up/down
in context $ \theta $, and $ N = n_+ +  n_- \rightarrow \infty $,  the total
number of electrons. Although the QM formalism does not itself attach to its
state functions a physical interpretation, in use $ \Psi $ is clearly to be
associated with sub-ensembles, in line with its realist interpretation
\footnote{ For an alternative and unconventional view see \cite{124}. With
regard to the special-case arguments put forward there, one might also
consider the special case of homogeneous ensembles.}.  An expansion of the
QM $\Psi $ over hypothetical realist electron states $ \psi $ might take the
form
\bea
|\Psi \ran &=& ( n_+  |+\ran_\theta +n_- |-\ran_\theta )/N \nonumber \\
&=& ( n_1  |\bs_1 \ran +n_2 |\bs_2 \ran )/N \label{Psi}
\eea
with
\bea
| \bs_i \ran &=& \sum_{ \{ \nu({\rm P }_i  ) = 1 \}} | \psi [ \{ \nu({\rm P
}^\prime ) \}^{\prime \prime} ] \ran \label{h3}
\eea
normalized to 1. Such an expansion leads immediately from the QM prediction
(\ref{h1}) to the observed (\ref{h2}).  The terms averaged over, $ \{\nu
({\rm P }^\prime ) \}^{\prime \prime} $, with their contextual dependence,
may be understood as the sub-ensemble's $ \bB $-context hidden variables.

\noindent As $ \hat{\bB} $ context eigenstates $ | \bs_i \ran $ will
generally not have definite values in other contexts, we have
\bea
0 \leq & \lan \bs_j^\prime | {\rm P }_i  | \bs_j^\prime \ran  &  \leq 1
\eea
although for realist possessed values
\bea
\nu^\prime ({\rm P }_i ) &=& 1\, {\rm or } \, 0 \, \qquad \forall \quad
\bB^\prime.    \label{h5}
\eea
A QM state function $ |\Psi \ran $ of (\ref{Psi} ) on the other hand has
definite values in at most one context and is therefore said in the realist
view to give a physically incomplete picture of the individual electron.
This difference forms the basis of our reconsideration of the KS paradox in
section 7.

\section{Kochen-Specker paradox }

\subsection{derivation of the paradox}

In line with the discussion in section 5 the following might be proposed as
a realist proposition
\begin{quotation}
\noindent p(1):  All real state quantities (spin, mass, etc.) have definite
values at all times.
\end{quotation}
which is known as the principle of value-definiteness. Pairing with this a 
second
proposition
\begin{quotation}
\noindent p(2):  The value of a real quantity does not depend on how it is
measured
\end{quotation}
called noncontextuality, and applying them together to the csr's (\ref{real 
eqns})
draws a contradiction in any Hilbert space of greater than two dimensions. 
As
the Hilbert space dimension for a particle of spin "s" is given by 2s+1, KS
shows that it is not possible to apply both p(1) and p(2) consistently for a
particle of spin $ s \geq 1 $; this is the KS
paradox \cite{106, 107}.

\noindent Let us consider the simultaneous application of the propositions
on (\ref{real eqns}).
From p(1) all equations and constraints hold simultaneously at this time
\begin{enumerate}
\item $ \nu^\prime ({\rm P }) = 0 \, {\rm or } \, 1 $
\item  for all $ \bB^\prime $
\end{enumerate}
and from p(2) the values are independent of context:
\begin{enumerate}
\item $  \nu^\prime ({\rm P } ) =   \nu^{\prime \prime  }({\rm P } ) \to
\nu ({\rm P } ) $.
\end{enumerate}
In their original work the authors next proceed with a lengthy formal 
argument, a simplification of which is the following :

\noindent The numbers $  \nu ({\rm P } ) $ are the eigenvalues of  Hilbert 
space
vectors $ | \bs_j \ran  $. According to the realist csr's (\ref{real eqns})
it must then be possible to assign to each mutually orthogonal set of three
vectors the values 0, 0, and 1. As the Hilbert space of QM state vectors,
H3, is complex, a violation of this rule in a Real space of the same
dimension will suffice for the negative proof of interest (R3 being a
subset of H3). Note however that the corresponding real vectors in the
mapping $ QM: R3 \to H3 \supseteq R3 $ (analogous to the 2-dimentional case
pictured in figure 7) do not coincide.

\noindent To begin, KS show in the following
way that two vectors separated by an angle
\bea
\phi  &\leq & \cos^{-1}(\surd8/3) \label{angle}
\eea
must have the same eigenvalue assignment, 0 or 1. Consider the three sets of
mutually orthogonal triplets
\bea
\{ |\bs_1^{(1)}>, |\bs_2^{(1)}>, |\bs_3^{(1)}> \} , & \{ |\bs_1^{(2)}>,
|\bs_2^{(2)}>, |\bs_3^{(2)}> \} , & \{ |\bs_1^{(3)}>, |\bs_2^{(3)}>,
|\bs_3^{(3)}> \} \label{nnnnew}
\eea
in R3 together with a tenth, $ |\bs_0> $, defined as
\bea
|\bs_0 \quad > & =& -x y \bi +x \bj -\bk   \nonumber \\
|\bs_1^{(1)}> &=& \bj+x \bk \nonumber \\
|\bs_2^{(1)}> & =& \bi    \nonumber \\
|\bs_3^{(1)}> & =& -x \bj +\bk   \nonumber \\
|\bs_1^{(2)}> & =& y \bi - \bj   \nonumber \\
|\bs_2^{(2)}> & =&  \bi + y \bj   \nonumber \\
|\bs_3^{(2)}> & =& \bk   \nonumber \\
|\bs_1^{(3)}> & =& - \bi -y \bj -x y\bk   \nonumber \\
|\bs_2^{(3)}> & =&   - y (1+x^2) \bi +(1-x^2 y^2 ) \bj + x (1+y^2) \bk
\nonumber \\
|\bs_3^{(3)}> & =&   -x y^3 (1+x^2) \bi + x (1+ 2 y^2 +x^2 y^2 ) \bj -
(1+y^2 )  \bk \label{mnew}
\eea
for arbitrary numbers  x  \&  y. Now the angle $ \phi $ between $ |\bs_0> $
and $ |\bs_3^{(3)}> $  found from their scalar product is given by relation
\bea
\cos(\phi ) &=& [1+x^2+y^2+x^2 y^2 (2+x^2+y^2+x^2 y^2) ] / ( ||\bs_0>||
||\bs_3^{(3)}>|| )
\eea
whose rhs minimization is directly found at $ \surd8/3 $, obtained for  $ x 
= y = \pm 1 $.
Therefore, $ 0 \leq \phi \leq \cos^{-1}(\surd8/3) $. In addition to the
explicit vector triplet constraints (\ref{real eqns}), by the same
consequence, no perpendicular pair of vectors may be assigned number "1",
since taken together with a mutually perpendicular third they would form a
triplet. One finds then from the additional relations
\bea
|\bs_0 \quad > & \perp & |\bs_1^{(1)}>, |\bs_2^{(2)}>, {\rm and } 
|\bs_2^{(3)}>
\nonumber \\
|\bs_1^{(3)}> & \perp & |\bs_3^{(1)}> { \rm and } |\bs_1^{(2)}> \nonumber \\
|\bs_2^{(1)}> & \perp & |\bs_3^{(2)}> \label{nnnew}
\eea
that the only possible assignments for the set of ten are those for which
$|\bs_0> $ and $ |\bs_3^{(3)}> $ have the same eigenvalue, $ \nu({\rm P }_0)
= \nu({\rm P }_3^{(3)}) = 0 \, {\rm or } \, 1 $, demonstrated by simply 
trying out each of the other two possible assignments while respecting constraints 
implicit in
(\ref{nnnnew}) and (\ref{nnnew}). As only the relative  $|\bs_0> $ to $
|\bs_3^{(3)}> $ directions are relevant, the result establishes for
KS that in realist thinking any two vectors in R3 of angular separation $ 0
\leq \phi \leq \cos^{-1}(\surd8/3) $ must have the same eigenvalue
assignment, 0 or 1.

\noindent We now take the full set $ \{|\bs_i^{(j)}> \}$ of (\ref{mnew}) for
$ \phi = 18^o ( < \cos^{-1}(\surd8/3) )  $, and for convenience perform a
rotation on the coordinate system such that $ |\bs_2^{(3)}> $ points along
the positive y-axis and $ |\bs_0> $ along the positive z-axis. We show here
three key vectors of the set scaled to 100, and table triplets according to
context $\hat{\bB}_i $.
\begin{center}
\scalebox{1.5}[1.5]{\includegraphics{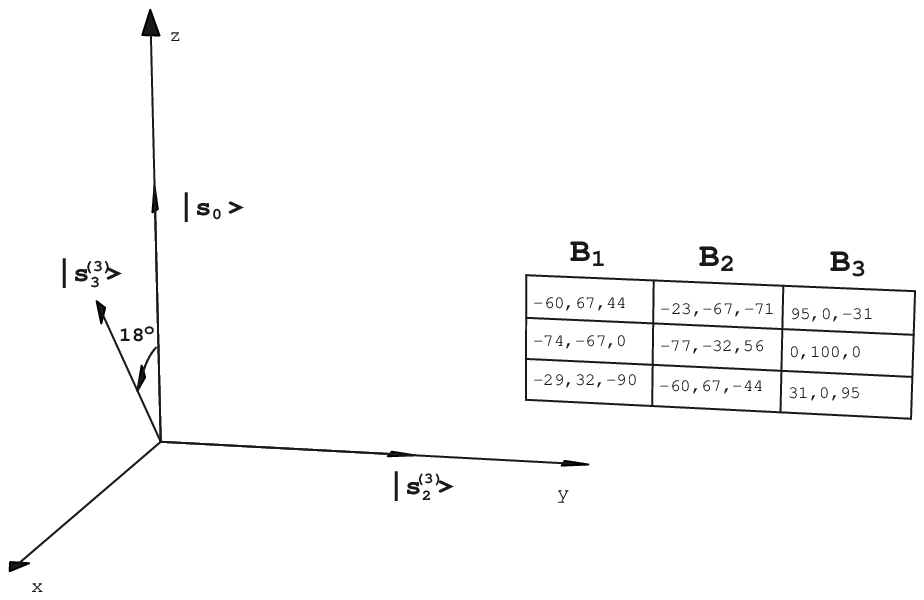}}\\
figure 11
\end{center}
From these we generate four new sets of nine vectors by rotating the
original about $ |\bs_2^{(3)}> $ in four $ 18^o $ increments, constraining 
successive
$|\bs_3^{(3)}> $'s thereby to the same value assignment, and
generate five additional sets by first rotating the last set by $ 90^o $
about its $ |\bs_3^{(15)}> $, then rotating in five more $ 18^o $ increments
about the rotated $ |\bs_2^{(15)}> $ . For a final five we follow the same
procedure to complete the first-quadrant sweep of $ |\bs_3 > $:  $
|\bs_3^{(15)}> ( = |\bs_2^{(33)}> ) \to |\bs_3^{(30)}> ( = |\bs_2^{(3)}> )
\to |\bs_3^{(45)}> ( = |\bs_2^{(18)}> ) $ ,  on the $ \{
\hat{\bi},\hat{\bj},\hat{\bk} \} $ triad
\begin{center}
\scalebox{1.5}[1.5]{\includegraphics{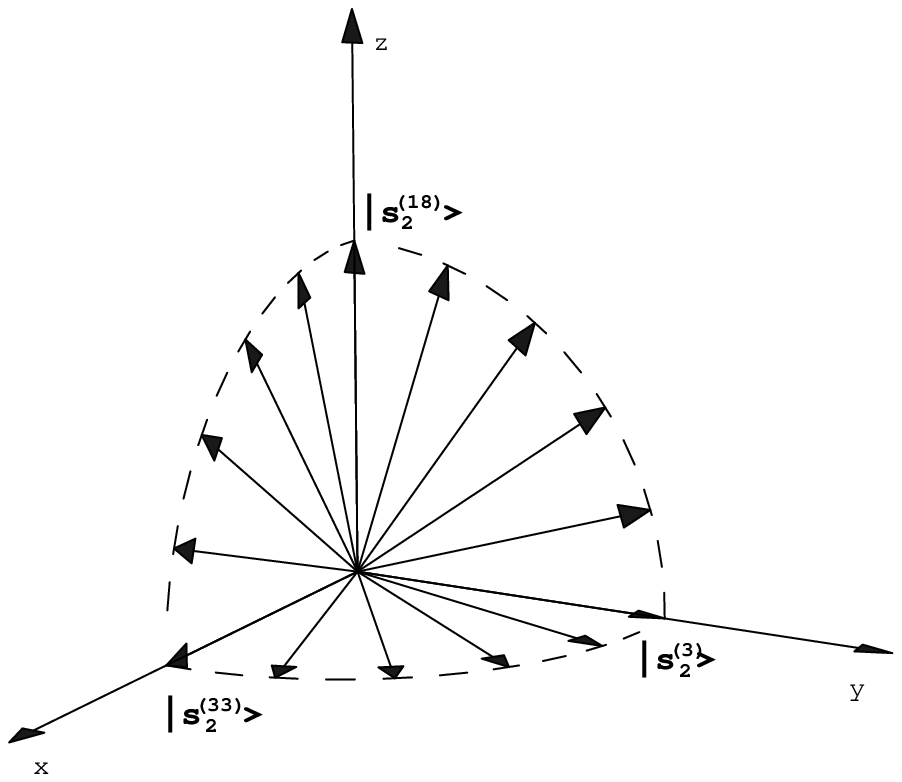}}\\
figure 12
\end{center}
each element of which, again, constrained to the same value assignment. But
this immediately violates csr's (\ref{real eqns}) as $ \{
\hat{\bi},\hat{\bj},\hat{\bk} \} $ is itself a mutually perpendicular
triplet. The violation is taken to demonstrate an inconsistency in the
realist conception of reality as given by propositions p(1) and p(2) and 
thus concludes
the KS paradox.

\noindent This result may also be illustrated visually. For the tabular form 
\cite{107} consider first the vectors involved in the 15
orientations of the set of 9 as represented in figure (12); we list here the
total 117 vectors (  = $ 15 \times 9 - 15 - 3$, accounting also for the 15 $
|\bs_2> $ and 3 $ |\bs_3>$ overlaps ), sub-grouping as in figure (11)
according to context
\begin{center}
\scalebox{0.9}[0.9]{\includegraphics{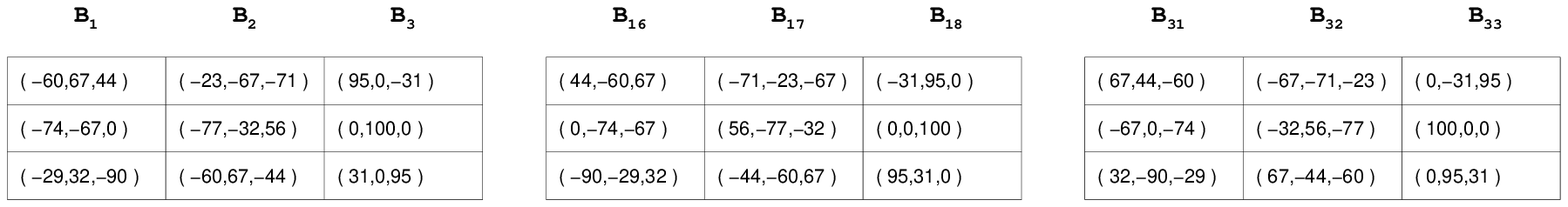}}\\
\end{center}
\begin{center}
\scalebox{0.9}[0.9]{\includegraphics{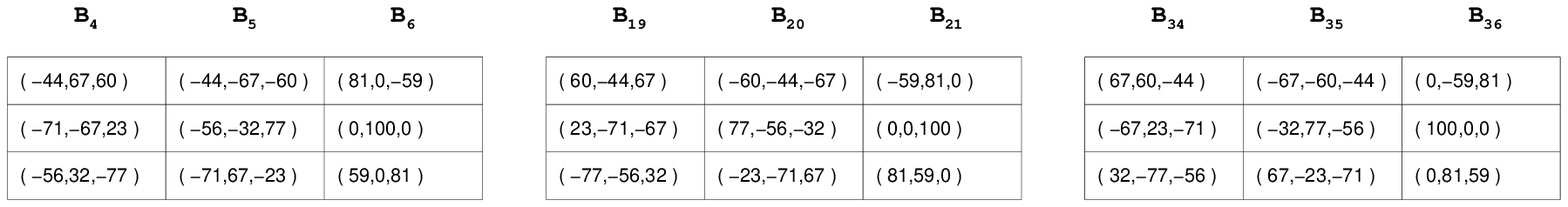}}\\
\end{center}
\begin{center}
\scalebox{0.9}[0.9]{\includegraphics{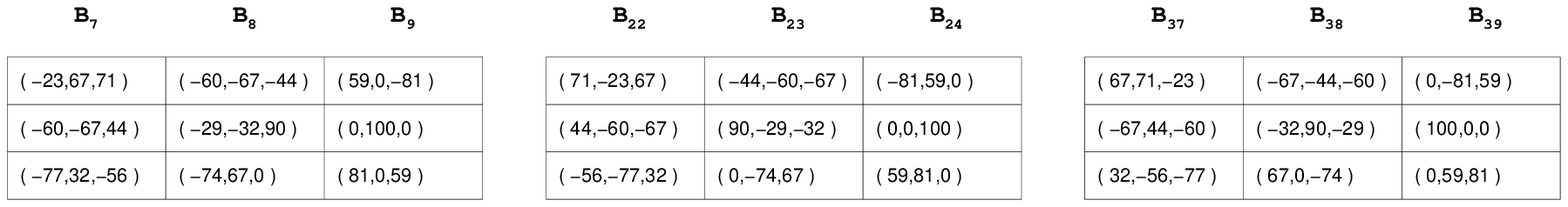}}\\
\end{center}
\begin{center}
\scalebox{0.9}[0.9]{\includegraphics{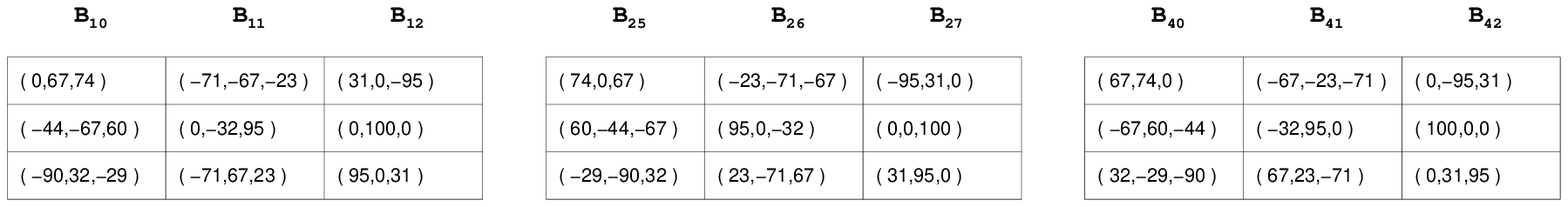}}\\
\end{center}
\begin{center}
\scalebox{0.9}[0.9]{\includegraphics{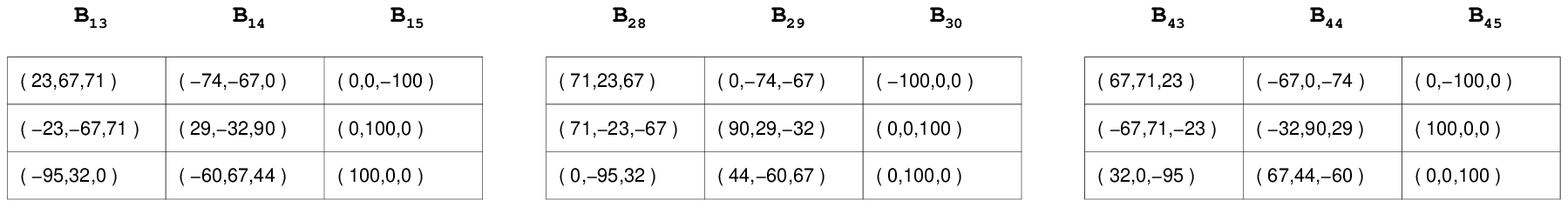}}\\
table 4
\end{center}
Then from the following rules
\begin{enumerate}
\item The eigenvector whose projector has value 1(0) is assigned the color
white (black).
\item Two orthogonal vectors may not both be colored white.
\item The assignments are independent of context.
\end{enumerate}
we attempt to color the vector boxes of table 4. That the coloring
assignments cannot be made consistently in accordance with the rules
coincides with the paradox.

\noindent Finally, for a two-dimensional graphical form of the paradox 
\cite{108} we represent vectors
by small open circles, and orthogonality between pairs of vectors by a
connecting straight line;  a triangle with open circles at the vertices then
represents a mutually orthogonal triplet, and our beginning set of vectors
described in figure (11) has the KS diagram representation
\begin{center}
\scalebox{0.9}[0.9]{\includegraphics{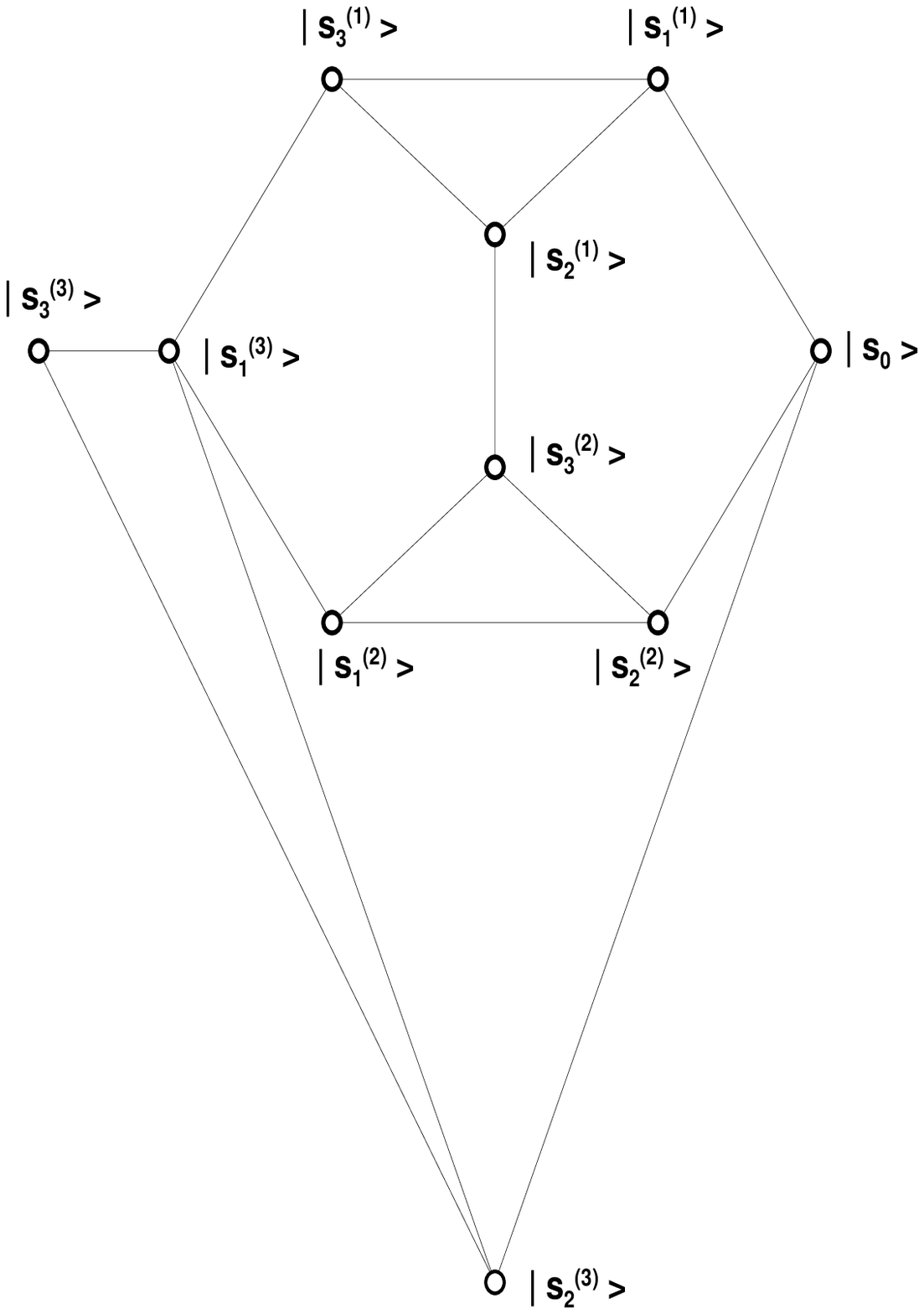}}\\
figure 13
\end{center}
Here the constraint that $|\bs_0> $ and $ |\bs_3^{(3)}> $ have the same
eigenvalue assignment for example follows from the inability to assign 
different colors
to their circled vertices in accordance with the aforementioned coloring
scheme for the corresponding vector boxes. If we then stretch the diagram a
little to one side
\begin{center}
\scalebox{0.9}[0.9]{\includegraphics{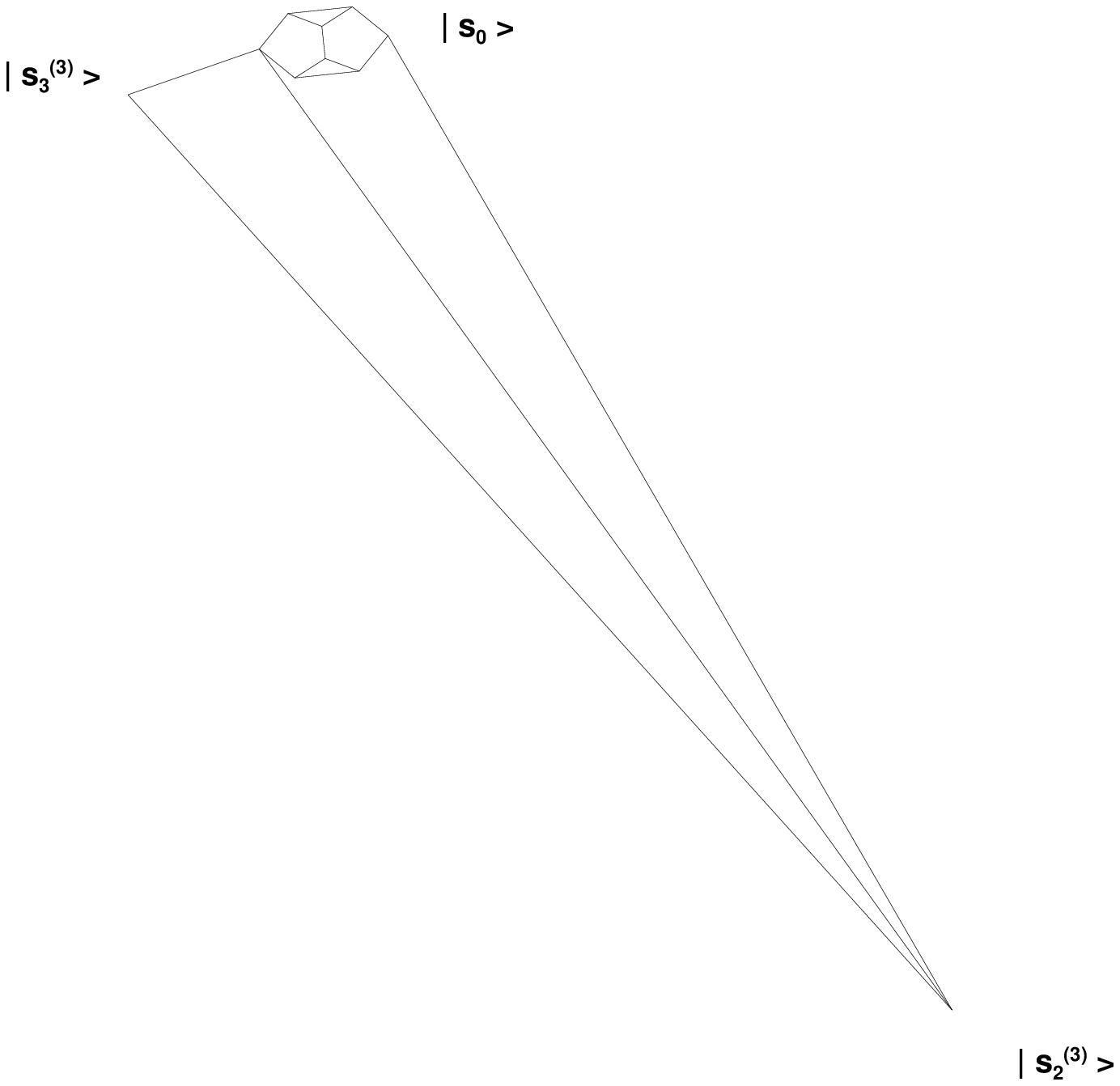}}\\
figure 14
\end{center}
and perform the 14 rotations about the appropriate $ |\bs_2> $ points (along 
with the two rotations about $ |\bs_3> $ points) as represented in
figure 12, we generate the full KS diagram
\begin{center}
\scalebox{0.9}[0.9]{\includegraphics{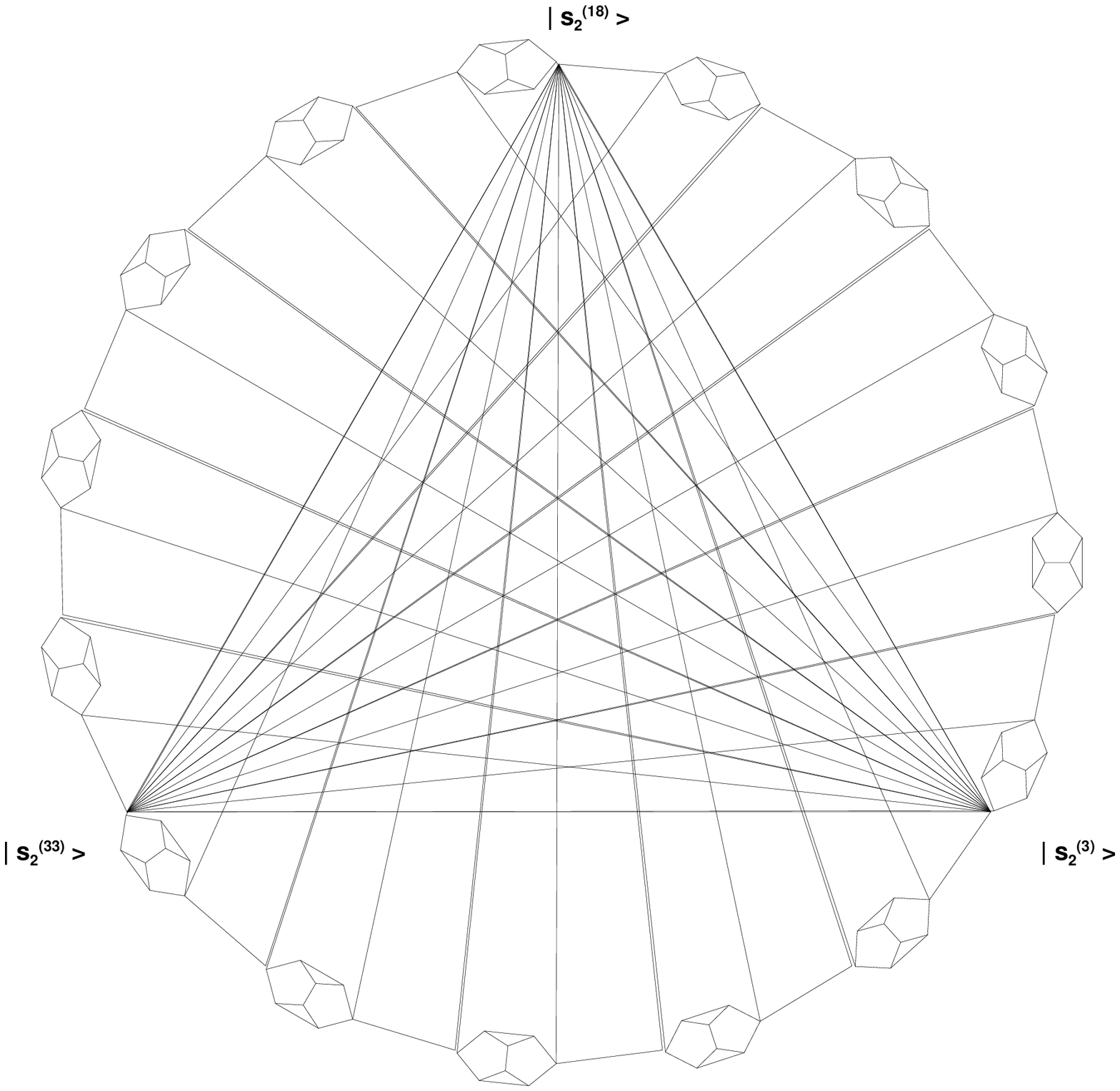}}\\
figure 15
\end{center}
said to resemble a cat's cradle.
Here too, the inability to faithfully apply the coloring scheme to
individual circled vertices is understood as a manifestation of inconsistency in tealist thinking on microscopic measurement. The difficulties presented by these two 
visualizations
are known as the KS {\it coloring problem}.

\subsection{Von Neumann's paradox and resolution}

In the Summer of 1932 eminent physicist and mathematician J. von Neumann
devised a proof (VN) against the possibility of a consistent realist
interpretation of QM \cite{109}, a precursor of the KS paradox, by way of the operator avarage relation
\bea
\bA = \bB + \bC  & \Rightarrow & \lan \Psi | \bA|\Psi \ran = \lan \Psi |
\bB|\Psi \ran + \lan \Psi | \bC | \Psi \ran
\eea
similar to our equation (\ref{h0}) above.
Operators $\bB$ and $\bC $ here do not commute, and the rhs terms are
assumed to take on realist possessed values \cite[p. \ 268]{104}. This last
condition is the VN additivity postulate P.IV from which derives the
expectation value formula
\bea
\lan O \ran &=& {\rm Tr } (W O) \label{trace}
\eea
where Tr is the trace operator and W a statistical operator independent of
observable O and descriptive of the system under observation. For the hypothetical value-definite
states $ \psi $, so-called "dispersion -free" states   
\bea
\lan \psi | O^2 | \psi \ran - \lan \psi | O | \psi\ran^2 &=& 0 \label{disp}
\eea
in the language of the time, the formula
proves problematic in the following way: The statistical operator describing the
realist system $ \psi $ is just the projector $ W_\psi = | \psi \ran \lan
\psi | $. Choosing for O then an arbitrary projector $ |\phi \ran \lan \phi
|  $ leads from (\ref{trace}) and (\ref{disp}) to the statement
\bea
\lan \phi | {\rm W }_\psi  | \phi \ran &=& 0 \, {\rm or } \, 1  \label{stat}
\eea
for all possible QM states $\phi $ which cannot be true as the left-hand
side is continuous in $ \phi $ and the right hand side discontinuous. This cast doubts on the very existence of dispersion-free
states $ {\rm W }_\psi $ \footnote{ I don't know that its been pointed out
in the literature \cite{114} that the above lhs continuity is
consequent upon a separate application of additivity by which two
dispersion-free states of opposite eigenvalues, $ \psi_1 \, {\rm and } \,
\psi_0 $, are joined by a connecting vector $ \Psi(x) = a_x \psi_0 + b_x
\psi_1 $, for properly chosen functions $ a_x $ and $ b_x $ ( See e.g.
reference \cite[p. \ 294]{104} ). Such an operator,  $ O =  |\Psi \ran \lan
\Psi |  $ is not an observable in the Hilbert space of dispersion-free
states under consideration. In light of this, the von Neumann proof might be
understood to follow from an assumption, under additivity, of the existence
of dispersive states in, under realism, the Hilbert space of dispersive-free
states: $ \lan \psi | \Psi \ran = 0 $ or $ 1 $ for all $ \Psi $. }.

\noindent This VN impossibility proof has long been understood to follow from a
misapplication of its additivity postulate \footnote{ See J. Bell,
ref.\cite{111}. A more thorough exposition is given in \cite[pp. \
25-34]{116} } which is experimentally valid for expectation values, though, as shown
in section 5.2, not applicable to real-possessed values. The violation follows from simple statistics:
ensemble values ($ \sim \Psi^2  $ ) are generally not identical with
ensemble-member values ( $\sim \psi^2 $).
The Von Neumann paradox rests upon an illicit application of proposition
p(1) to quantum mechanical sub-ensembles via equations of QM which $\Psi $
alone satisfies:
\begin{center}
$\Psi $ + p(1) $ \Rightarrow $ VN Paradox
\end{center}
( $\Psi $ implying p(2) ). It has the schematic resolution
\begin{center}
$ \Psi  \rightarrow  \psi $
\end{center}
or equally well,
\begin{center}
p(1) $ \rightarrow $ p(2).
\end{center}
We have then
\begin{equation}
\left. \begin{array}{c} \psi  + p(1) \\ {\rm or } \\ \Psi + p(2) \end{array}
  \right\} \Rightarrow  \qquad {\rm no \; VN \; Paradox }
\end{equation}
\subsection{Gleason's result }

In response to the objectionable additivity postulate necessary to the VN
proof, A. M. Gleason derived the main there formula (\ref{trace}) by means of an
acceptable additivity for commuting operators \cite{113}, as appears e.g. in
the above csr's (\ref{real eqns}), in addition to a tacit assumption of
noncontextuality first noticed only years later by Bell \cite{111}. By
demonstrating via (\ref{trace}) the necessary continuity of mappings from H3
and higher dimensions to R1, i.e. that Hilbert space projection operators ${\rm P
}_i $ are mapped continuously onto the closed interval [1,0],  Gleason
proves under noncontextuality the impossibility of the realist necessarily
discontinuous mapping $\nu : {\rm P } \to R1 , \,\, \nu( {\rm P }) = 0 \,
{\rm or }\, 1 $ , and so the non-existence of $\psi $ and an inconsistency
in realist hidden variable theories. The key result of Kochen and Specker,
constraint (\ref{angle}) , then gives an upper-bound estimate on the $
\nu^\prime(\phi) $ implicit to noncontextual value-definite hidden variable
theories.

\section{the KS paradox reconsidered }

The KS proof improves on the VN proof in that it in effect applies the
constraints of value definiteness, proposition p(1), appropriately to the
hypothetical realist state function $ \psi $ ($ \sim \nu $ ). Proposition
p(2), noncontextuality, on the other hand has largely been inferred
indirectly as an extension to realist thinking. For example, from Einstein's rhetorical
\begin{quotation}
Do you believe the moon is really there if no one is looking? \cite{119} (e)
\end{quotation}
which suggests an independence of reality from its observation.

\noindent Let us briefly take a closer look. As the statement addresses the question of existence itself, the observation-independence or
noncontextuality inferred from it is ontological\footnote{ Ontological in the sense that it concerns the principles
and causes of being.... Unfortunately for the writer, the term
"ontological contextualism" is already in use \cite{117} though with a meaning
it seems closer to that of causal contextuality. I use it here sparingly
in the first sense. }: There is an objective reality "out there"
with an {\it existence } independent of observation \footnote{ See e.g. K.
Popper in ref.\cite[p. \ 2]{112}  }. The aforementioned reality criterion of
section 5.1, however concerns an observation
dependence of a different, more mundane sort: The observation-dependence of
states of existence, of {\it states } (which addresses e.g. the question whether
an electron's spin is affected by an observation of it ), not of the very
existence of the thing observed. The clear indication is that,
notwithstanding the select elements that meet the reality criterion - a
sufficient criterion - the value of a physical quantity does indeed depend
upon how it is obtained, upon the manner in which the object under
observation is affected, disturbed during measurement. Such a contextual
dependence is causal and the values thus measured causally contextual - precisely the sort of effect that experimentalists in EPR tests go to great lengths to guard against \cite{110}. As J. Bell writes,
"The result of an observation may reasonably depend not only on the state of
the system... but also on the complete  disposition of the [measuring]
apparatus." \cite{111}. But these categories are confused in the KS and Gleason
analysis, realist dismissal of ontological contextuality misinterpreted as a
dismissal of causal contextuality, leading, finally, to mistaking p(2) as a realist
proposition.

\noindent While the von Neumann proof relies on an explicit misapplication
of realist value-definiteness within the QM formalism, as shown in 5.1, in its
misapplication of additivity via the equations of QM such as (\ref{d1})
to realist states $ \psi $, it is also seen to violate causal contextuality; in the single QM expectation-value equation
(\ref{h1}) the relevant contextual effects are washed out in the average. KS
and Gleason achieve the same ends by means of properly single
context equations, though several taken simultaneously in terms of the csr's of (\ref{real
eqns}) for mutually exclusive contexts. And so in principle both commit the same violation in unduly
constraining the most general causality. Again, to illustrate, there is
nothing in realist thinking on the spin-1 measurements
\begin{center}
\scalebox{1.5}[1.5]{\includegraphics{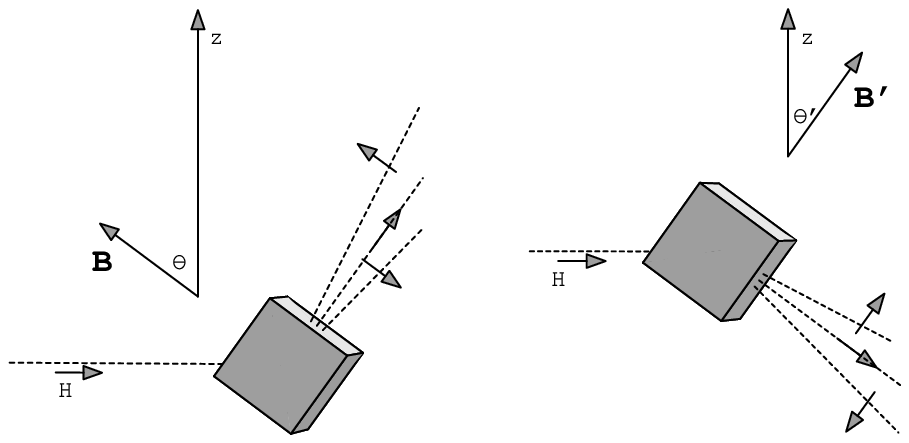}}\\
figure 16
\end{center}
($\theta - \theta^\prime = 90^o $) to suggest the counterfactual implied in
the noncontextual constraint $\nu (P_1) = \nu^\prime (P_2) $ ( or, with
reference to figure 10, the constraint $\nu (P_2) = \nu^\prime (P_2) $ ).

\noindent The KS diagram of figure (15) describes a set of experiments each
of which may in principle be performed; accordingly, the diagram itself may
always be constructed. Its coloring however depends somewhat upon interpretation of QM wavefunction.
While both orthodox and realist interpretations achieve coloring consistency
by means of various shadings, indeed in violation of the coloring rules,
their mechanisms differ: Indeterminism on one hand, causal contextuality on
the other. At issue here is the KS contention that while orthodox coloring
is constrained only by the rules governing individual triplets, those that
might be derived from the csr (\ref{new}), realist coloring is constrained
by the full KS set.

\noindent Proposition p(2) from which in combination with p(1) the rules
follow does not properly apply to individual ensemble-member values, realist
possessed values $ \nu $,  and it is precisely this misapplication that is at the heart of the KS paradox.

\noindent The paradox may thus be understood from the following schematic:
\begin{center}
$ \psi $  + p(2) $ \Rightarrow $ KS Paradox
\end{center}
($ \psi $ implying p(1) ). It is resolved by the replacement
\begin{center}
$ \psi   \rightarrow  \Psi $
\end{center}
Taken together then
\begin{equation}
\left. \begin{array}{c} \Psi  + p(2) \\ {\rm or } \\ \psi  + p(1)
\end{array}  \right\} \Rightarrow  \qquad {\rm no \; KS \; Paradox }
\end{equation}

\section{conclusion}

While quantum mechanics is generally accepted as the best tool available for
the prediction of small-scale statistics, there remains disagreement over
its physical interpretation particularly as regards its
Hilbert-space state vectors. In the prevailing orthodox view $ \Psi $ is
said to represent individual systems e.g. an individual electron of an
ensemble.  It is understood in consequence that individual systems cannot
consistently possess many of the common characteristics we
attribute to them in ordinary language use (spatial position, mass, spin,
etc.). The conceptual challenges presented by the view are obvious and
well-known and an entire culture in science thinking has gone  to their
explication \cite{112,104}. While this particular concept does indeed appear
common to all orthodox-school thinking we caution not to overgeneralize
with regard what is really a diversity of views \cite{140}.
\noindent Likewise in characterizing the realist view \footnote{
See refs.\cite[p. \ 44]{108}  and \cite[p. \ 9]{116}  } , particularly as
impossibility proofs such as the ones here reviewed are directed against their
very logic. Let us therefore carefully consider again the type of experimental
arrangement cited earlier in illustration of realist thinking on the fundamental
issue here, noncontextuality.

\noindent For the pair of illustrated spin-1 projection measurements
\begin{center}
\scalebox{1.5}[1.5]{\includegraphics{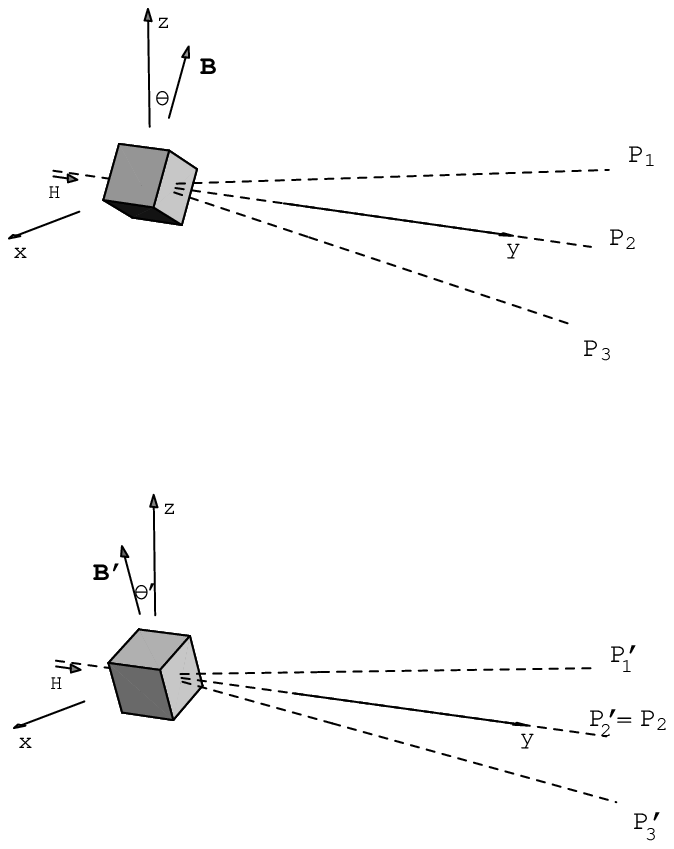}}\\
figure 17
\end{center}
one always has the possessed-value relations
\bea
\nu({\rm P }_1) + \nu({\rm P }_2) + \nu({\rm P }_3) &=& 1
\eea
and
\bea
\nu^\prime ({\rm P }_1^\prime) + \nu^\prime ({\rm P }_2^\prime) + \nu^\prime
({\rm P }_3^\prime) &=& 1
\eea
We ask now whether in the realist view a result of measuring the projection $
{\rm P }_2 = {\rm P }^\prime_2 $ may depend upon how the measurement is taken. Whether
\bea
\nu ({\rm P }_2) &=& \nu^\prime ({\rm P }_2^\prime) \,  \label{1?}
\eea
by necessity.
Based on statements of the generality (e) above KS assumes that the realist
view answers in the affirmative, the claim (not to beat a dead
horse) at the heart of the derivation of their paradox, as shown in section
6.1.

\noindent But the question is not unique to experimental arrangements of the
type pictured in figure (17). The very same also appears in Bell -Inequality
(BI) analysis \cite{111} in the context of the question of space-time locality
\footnote{ There, it is certainly true that what is to the realist view a question
of contextuality (of the possible effects of $ {\rm D}_2 , \psi_2 , { \rm
etc. \, \, on } \, \, \psi_1 $ , {\it assuming separability} - that the
effects between paired particles are purely causal) is to the orthodox view
one of separability only ( roughly, of the affects of decay particle $
\psi_2 $ on particle $ \psi_1 $ intrinsic to their bound state ). Hence e.g.
the realist issue with detector $ {\rm D}_2 $ triggering. }. There, a set of
fixed projection measurements is taken (by detector $D_1$ at space-time
position x ) on spin-1/2 particles, as the setting of a remote apparatus
($D_2$ at y ) is varied
\begin{center}
\scalebox{1.5}[1.5]{\includegraphics{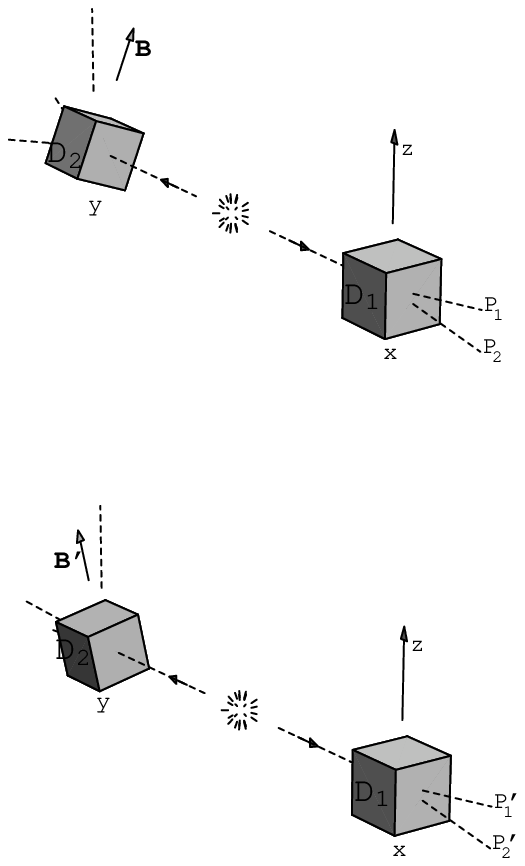}}\\
figure 18
\end{center}
One then considers contextuality question, whether
\bea
\nu_\bk ({\rm P }_1) &=& \nu_\bk^\prime ({\rm P }_1)  \label{2?}
\eea
where here too value subscripts and superscripts in obvious notation
characterize the measurement context \footnote{ While it is possible for
individual states and measurements $ (\psi, \; \nu ) $  to depend upon
context while those of ensembles $ ( \Psi, \lan \; \ran ) $ remain
independent, or noncontextual, Bell's inequality is concerned with ensemble
contextuality (that of $ \Psi $ ) though by way of individual state
contextuality (that of  $ \psi $ ). The KS paradox on the other hand treats the contextuality of individual states.}.

\noindent An affirmative answer without condition leads as is well-known
to an realization of the Bell inequality and immediate disagreement of
realism with observation. And no one has thought to do this. On the contrary,
contextuality is assumed as a matter of course, its possible effect dismissed only
upon satisfaction of the most stringent locality conditions
\bea
( x - y )^2 & > & 1 \label{3?}
\eea
brought to some level of experimental realization in the work of
Aspect et al and several refinements since \cite{110}. We are keen to
make this point, as (to continue the beating) causal contextuality is indeed
central to realist thinking particularly since the appearance of quantum
phenomena and the realist conceptual involvement of hidden
variables. D. Bohm confirms that " when we measure the momentum
'observable,' the final result is determined by hidden parameters in the
momentum-measuring device as well as by hidden parameters in the observed
electron" \cite{120}. N. Bohr also, himself no realist, agrees \footnote{
"This crucial point... implies the impossibility of any sharp separation
between the behavior of atomic objects and the interaction with the
measuring instruments which serve to define the conditions under which the
phenomena appear.", Ref. \cite[p. \ 209]{121} }, while others, more often
than not on purely formalistic grounds, may not \cite{123}.  We conclude therefore that
the value-definiteness of realism p(1) is in no way incompatible with causal
contextuality not-p(2) but is on the contrary complemented by it \footnote{ Without naming it, as the term had yet
to enter the language, Belinfante offers in part 1 of reference \cite{116}
(up to page 78) an excellent insight into contextual thinking. Likewise Bell
\cite{111}, the first to notice contextuality as an issue central
to the KS paradox. There are many others.}.

\noindent If figure(18) by comparison with (17) (in light of constraint
(\ref{3?})) is taken to indicate an experimental constraint appropriate to noncontextuality in the KS analysis, the feasibility of such a test seems
clearly in doubt. We will not persue the question here.

\noindent What is striking about the realist view of possessed values is its
marked lack of novelty. The possessed values of an electron are thought to
be no different in their manner of possession than tose of tables, trees,
cats; a stone said to possess 10 kilograms is said in the same breath to
weigh 98 Newtons, to oscillate with a frequency $ f =  \frac{1}{2\pi }
\surd\overline{\frac{k}{10}} $ \, when attached to a spring k, and so
forth; i.e., such a stone possesses all the values that a stone of 10
kilograms might yield upon measurement, many of which from the laws of mechanics may be known before hand.... To say that the stone is 10 kilograms is a manner of speaking, a shorthand for an otherwise
unwieldy usage.

\noindent In absence such a shorthand however one is left with the
directly observed or predicted values themselves, no
less possessed by the object. In this form they may well appear in some ways
peculiar and to have unusual properties \footnote{ van Fraassen's
ontological contextualism has taken special ridicule for its proliferation
of possessed values that e.g. "pop in and out of existence" at the whim of
the experimentalist \cite{107}, "de-Ockhamizing QM!" \cite[p. \ 135]{108} .
Such a criticism on mathematical form, Bell might have countered, merely
betrays a lack of sufficient imagination \cite[p. \ 64]{200}. Einstein too
was largely unconcerned with formal structures per se \cite[pp. \
234-238]{104} . } . They do not; and the appearance generally dissolves upon
careful consideration: Again, the manner of microscopic property possession in this view is no different than that for macroscopic possession.

\noindent It has long been considered not possible to assign definite values to
systems whose possessed values are described by noncommuting QM operators.
The KS paradox proves this for a discrete set of observables and as such
is a statement of the limitation of QM to predict individual microscopic phenomena. But the paradox is not a proper criticism
of realist thinking; it reduces to absurdity a view held by no one.


\begin{thebibliography}{}
\bibitem{114}
There seems to be, according to \cite{115}, some indication of this in J.
Tharrats, Comptes Rendus {\bf  250}, 3786 (1960), with which, by the way
Clifton in \cite{115} emphatically disagrees. But I have not found the
English translation.
\end{thebibliography}
\end{document}